\documentclass[superscriptaddress, preprint]{revtex4-1}

\usepackage{xspace}
\usepackage{times}
\usepackage{graphicx}
\usepackage{amsmath}
\usepackage{stmaryrd}
\usepackage[
  linktocpage,
  linkbordercolor={0.5 0.5 1},
  citebordercolor={0.5 1 0.5},
  linkcolor=blue]{hyperref}

\newcommand{\rh}[1]{\ensuremath{\rho_{\mathrm{#1}}}\xspace}
\newcommand{\degr}[1]{\ensuremath{#1^\circ}\xspace}
\newcommand{\thh}[1]{\ensuremath{#1^{\mathrm{th}}\xspace}}
\newcommand{\bpaj}{\ensuremath{\mathrm{\textbf{J}}||\mathrm{\textbf{B}}||\mathrm{\textbf{c}}}\xspace}
\newcommand{\bpej}{\ensuremath{\mathrm{\textbf{J}}\!\!\perp\!\!(\mathrm{\textbf{B}}||\mathrm{\textbf{c}})}\xspace}

\begin{document} 

\title{Quantum Limit Transport and Destruction of the Weyl Nodes in TaAs } 

\author{B.~J.~Ramshaw}
\email{bradramshaw@cornell.edu}
\affiliation{Laboratory of Atomic and Solid State Physics, Cornell University, Ithaca, NY, 14853.}
\affiliation{Los Alamos National Laboratory, Los Alamos, New Mexico, 87545.}
\author{K.~A.~Modic}
\affiliation{Max-Planck-Institute for Chemical Physics of Solids, Dresden, Germany, 01187.}
\author{ Arkady Shekhter}
\affiliation{National High Magnetic Field Laboratory, Tallahassee, Florida, 32310.}
\author{ Yi Zhang}
\affiliation{Laboratory of Atomic and Solid State Physics, Cornell University, Ithaca, NY, 14853.}
\author{ Eun-Ah Kim}
\affiliation{Laboratory of Atomic and Solid State Physics, Cornell University, Ithaca, NY, 14853.}
\author{ Philip J.~W.~Moll}
\affiliation{Max-Planck-Institute for Chemical Physics of Solids, Dresden, Germany, 01187.}
\author{ Maja D. Bachmann}
\affiliation{Max-Planck-Institute for Chemical Physics of Solids, Dresden, Germany, 01187.}
\author{ M.~K.~Chan}
\affiliation{Los Alamos National Laboratory, Los Alamos, New Mexico, 87545.}
\author{ J.~B.~Betts}
\affiliation{Los Alamos National Laboratory, Los Alamos, New Mexico, 87545.}
\author{ F.~Balakirev}
\affiliation{Los Alamos National Laboratory, Los Alamos, New Mexico, 87545.}
\author{ A.~Migliori}
\affiliation{Los Alamos National Laboratory, Los Alamos, New Mexico, 87545.}
\author{N.~J.~Ghimire}
\affiliation{Argonne National Laboratory, Argonne, Illinois, 60439.}
\affiliation{Los Alamos National Laboratory, Los Alamos, New Mexico, 87545.}
\author{ E.~D.~Bauer}
\affiliation{Los Alamos National Laboratory, Los Alamos, New Mexico, 87545.}
\author{ F.~Ronning}
\affiliation{Los Alamos National Laboratory, Los Alamos, New Mexico, 87545.}
\author{ R.~D.~McDonald}
\affiliation{Los Alamos National Laboratory, Los Alamos, New Mexico, 87545.}

\maketitle

\section{Abstract}
Weyl fermions are a new ingredient for correlated states of electronic matter. A key difficulty has been that real materials also contain non-Weyl quasiparticles, and disentangling the experimental signatures has proven challenging. We use magnetic fields up to 95 tesla to drive the Weyl semimetal TaAs far into its quantum limit (QL), where only the purely chiral \thh{0} Landau levels (LLs) of the Weyl fermions are occupied. We find the electrical resistivity to be nearly independent of magnetic field up to 50 tesla: unusual for conventional metals but consistent with the chiral anomaly for Weyl fermions. Above 50 tesla we observe a two-order-of-magnitude increase in resistivity, indicating that a gap opens in the chiral LLs. Above 80 tesla we observe strong ultrasonic attenuation below 2 kelvin, suggesting a mesoscopically-textured state of matter. These results point the way to inducing new correlated states of matter in the QL of Weyl semimetals.

\section{Introduction}
\begin{figure}
\centering
\includegraphics[width=\columnwidth,clip=true,trim=00mm 00mm 00mm 00mm]{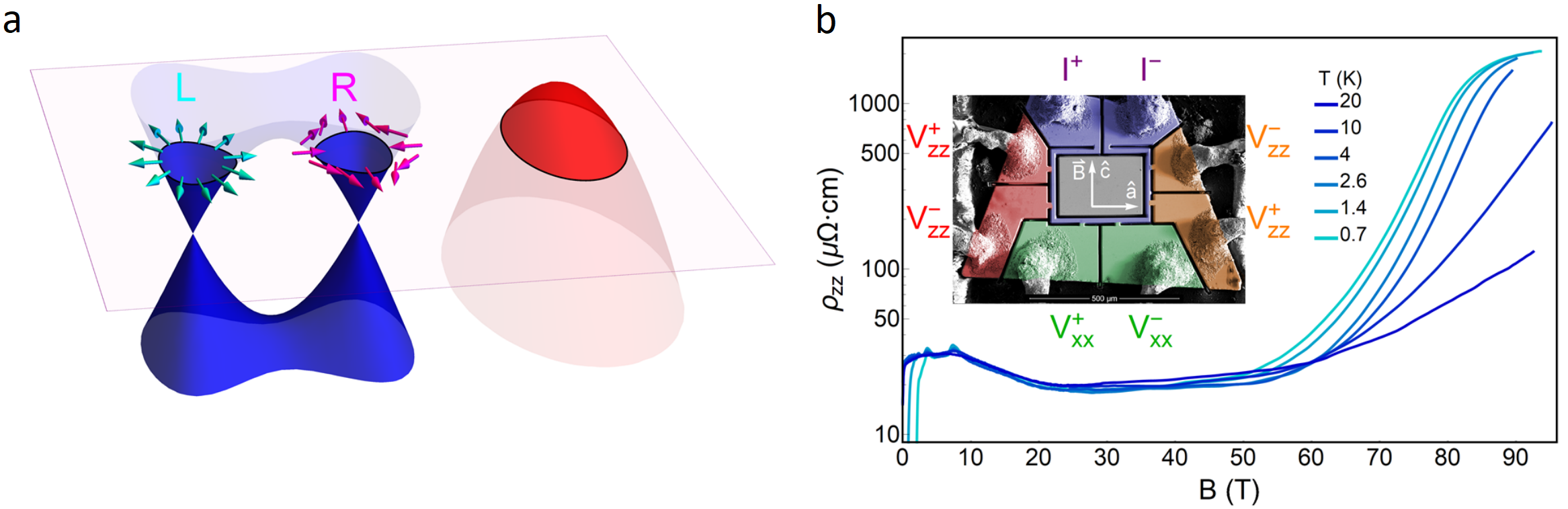}
\caption{\textbf{Weyl fermions and quantum limit transport in TaAs.} \textbf{a}, the Weyl semimetal TaAs contains three types of charge carriers: trivial holes (red), and two sets of electron-like Weyl fermions, designated W1 and W2, one of which is shown here in blue. The holes are of a single non-chiral carrier type, and contribute a non-Weyl background signature to experiments. The Weyl electrons are separated into distinct right and left-handed chiralities: arrows indicate the winding of the pseudospin around each Fermi surface. \textbf{b}, Resistivity of TaAs for \bpaj from 0.7 to 20 K. Quantum oscillations from the Weyl pockets are visible up to 7.5 tesla, followed by a decrease and then saturation of \rh{zz} up to 50 tesla. Above 50 tesla there is a two order-of-magnitude increase in \rh{zz} at low temperature, signifying the opening of a gap. The inset shows single-crystal TaAs microstructured using focused-ion-beam (FIB) lithography for both the \rh{zz} and \rh{xx} measurements.}
\label{fig:rzz}
\end{figure}

The principle of emergence is beautifully demonstrated by the complex states of matter that arise in metals due to interactions. Conventional metals can become superconductors \cite{Onnes:1911,Bednorz:1986}, spontaneously break spatial symmetries \cite{Kivelson:1998}, emerge with fractionalized charge \cite{Tsui:1982}, or even become insulating \cite{Mott:1968}. The recent prediction \cite{Wan:2011,Weng:2015} and discovery \cite{Xu:2015,Lv:2015} of Weyl semimetals brings a new ingredient to the table---Weyl fermions---that differ from conventional electrons in two key ways. First, they obey a linear or `massless' energy-momentum relationship, analogous to electrons in graphene but in three dimensions. Second, because of strong spin-orbit coupling and inversion symmetry breaking, spin `up' and 'down' states are replaced by chiral `left' and `right'-handed states (\autoref{fig:rzz}a) \cite{Nielsen:1983}. Can interacting Weyl fermions form states of matter that are unique from those that arise in conventional metals? Predictions include chiral excitonic \cite{Wei:2012} and density wave states \cite{Zhang:2017}, topologically non-trivial superconductors \cite{Cho:2012,Meng:2012}, and Tomonaga-Luttinger liquids \cite{Zhang:2017b}, but no interacting states have been confirmed in Weyl semimetals to date. Both Weyl fermions and trivial quasiparticles exist in real Weyl semimetals, making it difficult to tell which effects are due to Weyl fermions and which effects can be ascribed to conventional semimetal physics \cite{Armitage:2017}.

In addition to the possibility of new correlated states, Weyl semimetals exhibit a number of unusual non-interacting properties due to the chiral anomaly. Chirality, or `handedness', is a classically conserved quantity---the flows of left and right-handed quasiparticles (chiral currents) should be separately conserved. The correct quantum mechanical description of nature does not obey this symmetry, allowing for classically-forbidden phenomena such as the decay of a neutral pion into two photons \cite{Adler:1969,Bell:1969}. In the case of Weyl fermions it has been shown that parallel electric and magnetic fields can unbalance the number of left and right-handed Weyl fermions \cite{Nielsen:1983}, leading to exotic transport \cite{Nielsen:1983,Son:2013,Spivak:2016,Lucas:2016} and optical \cite{Hosur:2015} properties. While this effect was originally predicted for a Weyl semimetal far in the QL \cite{Nielsen:1983}, most experiments probing these effects have focused on relatively low magnetic fields \cite{Huang:2015,Lv:2015,Gooth:2017,Ma:2017}. Again, interpretation of these results is hampered by the fact that there are multiple Weyl fermions and non-chiral quasiparticles all present at the Fermi energy. By using high magnetic fields we are able to gap out the non-chiral quasiparticles in TaAs and move the chemical potential to the fully chiral $n=0$ LLs belonging to the Weyl nodes. This allows us to directly observe the chiral anomaly in the QL, measure the opening of a gap in the $n=0$ LLs once the inverse magnetic length becomes longer than the momentum-space separation of the Weyl nodes above 50 tesla, and discover the onset of a new state above 80 tesla.
\section{Results}

\subsection{High field transport in TaAs}

\autoref{fig:rzz}b shows the resistivity of the Weyl semimetal TaAs in magnetic fields up to 95 tesla. Immediately striking is the two-order-of-magnitude increase in resistance that onsets at 50 tesla for the current and field configuration \bpaj (note the logarithmic scale). In addition to its strong magnetic field dependence, \rh{zz} is also strongly temperature dependent above 50 tesla. This is in contrast with the behaviour at intermediate fields (between 7.5 and 50 tesla), where \rh{zz} decreases and then saturates, becoming roughly temperature and field independent. We identify the transport behaviour between 7.5 and 50 tesla as characteristic of the chiral anomaly in the QL for Weyl fermions, and the behaviour above 50 tesla as indicative of mixing between left and right handed Weyl fermions. There are two features in the data that we do not believe to be inherent to the bulk transport of TaAs: a superconducting transition at low field, originating in a $\approx 20$ nm tantalum-rich amorphous layer induced during sample preparation \cite{Bachmann:2017}; and the rollover of \rh{zz} at high field, either due to the same amorphous layer or due to the intrinsic surface states of TaAs \cite{Lv:2015,Lv:2015b,Zhang:2015}. Neither effect impacts the conclusions of this paper (see Methods).

\begin{figure} 
\centering
\includegraphics[width=\textwidth,clip=true,trim=0mm 0mm 0mm 0mm]{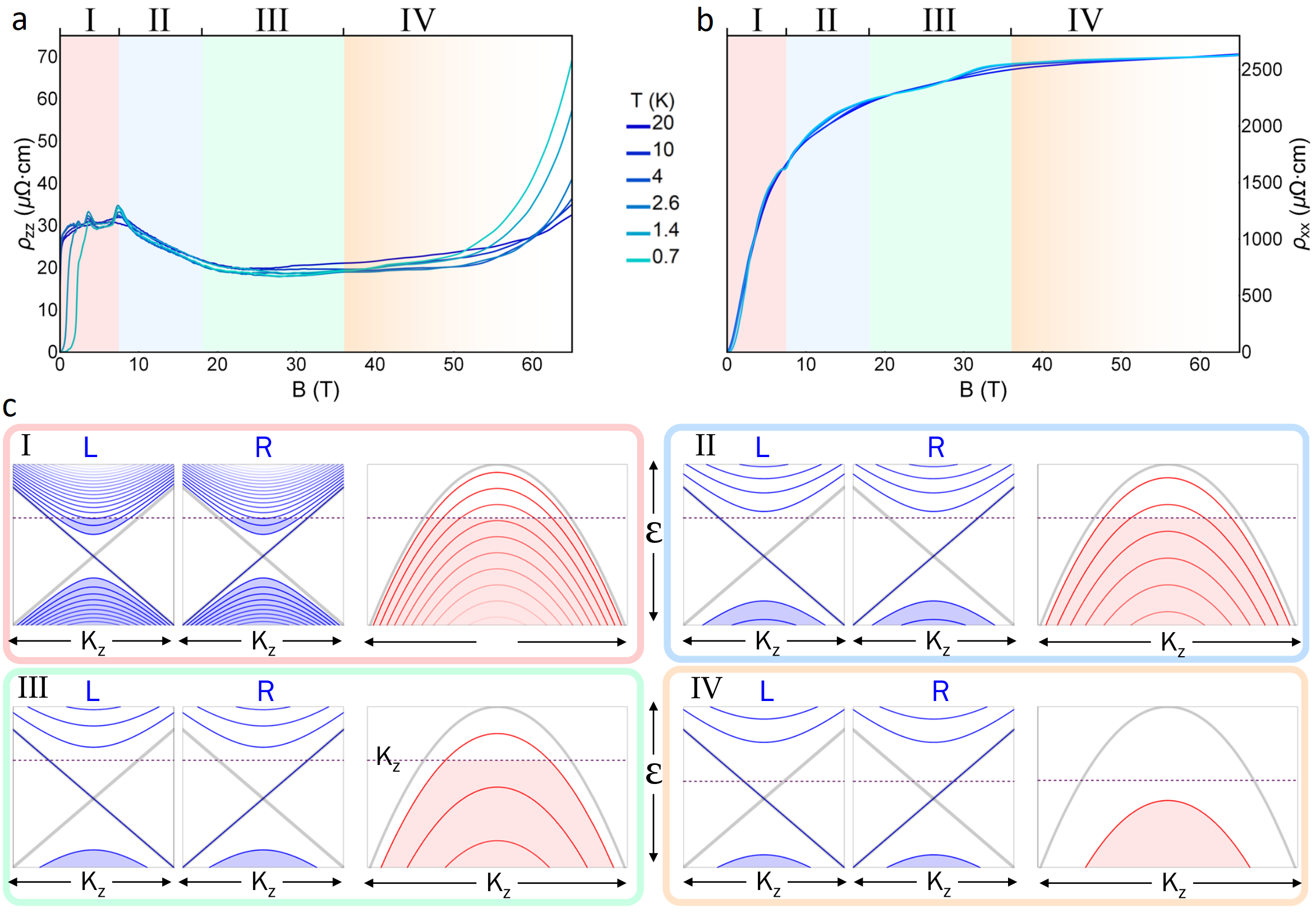}
\caption{\textbf{The four distinct regimes of Landau level occupancy in TaAs up to 65 tesla.} TaAs hosts two pairs of Weyl fermi surfaces (W1 and W2) with similar quantum limits for this field orientation; one pair of Weyl nodes is shown here along with the trivial hole surface. Resistivity of TaAs at temperatures between 0.7 K and 20 K, for \bpaj (\textbf{a}) and for \bpej (\textbf{b}), from 0 to 65 tesla. \textbf{c,} Schematic of the Landau level (LL) structure and occupancy for the parabolic hole pockets (red lines) and Weyl electron pockets (blue lines). Grey curves are the $B=0$ dispersions without LLs. Region \textbf{I} is from 0 to 7.5 tesla, where both hole and electron pockets exhibit quantum oscillations. Between 7.5 and 18 tesla (region \textbf{II}) the electron pockets are in the $n=0$ LL, whereas the hole pocket is still in the $n=1$ LL. Between 18 and 36 tesla (region \textbf{III}) the hole pocket is also in the $n=0$ LL. At 36 tesla the last oscillation, corresponding to emptying the $n=0$ LL of the hole pocket, can be seen in \rh{xx}. Above 36 tesla (region \textbf{IV}) the only occupied states at the chemical potential are in the $n=0$ LL of the Weyl electron pockets, and the chemical potential shifts to maintain overall carrier number ($n_e-n_h$). The faded region \textbf{IV} represents the mixing of the Weyl nodes above 50 tesla (\autoref{fig:mixed}). }
\label{fig:lls}
\end{figure}

We measure both \bpaj---\rh{zz}---and \bpej---\rh{xx}---to identify different regimes in the electronic structure of TaAs (Figure \ref{fig:lls}). TaAs contains three distinct types of quasiparticles: two electron-like Weyl Fermi surfaces (denoted W1 and W2 \cite{Lv:2015,Lv:2015b,Huang:2015,Arnold:2016,Ma:2017}), and a hole-like Fermi surface with a `trivial' spectrum \cite{Huang:2015,Zhang:2016,Arnold:2016}. \autoref{fig:lls}c highlights a key difference in the Landau quantization between Weyl and trivial carriers: the energy of the $n=0$ LL of Weyl carriers is field-independent and disperses linearly in $k_z$ (momentum parallel to $\mathrm{\textbf{B}}$), whereas the energy of the $n=0$ LL of trivial carriers is field-dependent and disperses quadratically in $k_z$. This leads to qualitatively different behaviour for the two carrier types in the QL, when only the $n=0$ LL is occupied \cite{Moll:2016}. As the magnetic field is increased TaAs passes through four distinct LL regimes, described in detail in the caption to \autoref{fig:lls}. With the magnetic field aligned along the c-axis both the W1 and W2 Weyl pockets enter the QL (the chemical potential enters the $0^{\mathrm{th}}$ LL) below 8 tesla, as observed in both our data and in previous studies of the Fermi surface of TaAs \cite{Arnold:2016}. The hole pocket enters the quantum limit at 18 tesla and is completely depopulated at 36 tesla---also agreeing with previous studies \cite{Arnold:2016}. The fact that the trivial hole carriers can be completely depopulated with field is a result of the field-independence of the $n=0$ LL of the Weyl pockets, where the chemical potential is pinned due to the small excess of electrons versus holes.

\subsection{Chiral anomaly in the quantum limit}
Above 36 tesla the chiral $n=0$ LLs of the Weyl fermions can be studied without the complicating influence of trivial holes, and without the $n>0$ LLs for which chirality is not well-defined \cite{Arnold:2016b}. Each of the two \thh{0} LLs of the Weyl fermions contains carriers propagating along a single direction---parallel to $\mathrm{\textbf{B}}$ for one chirality (`up movers'), and anti-parallel for the other (`down movers'). Applying an electric field (driving current) parallel to $\mathrm{\textbf{B}}$ produces an imbalance in the number of up versus down movers, resulting in an imbalance of left and right-handed carriers---a chiral anomaly. This imbalance produces a chiral current, effectively increasing the conductivity as the electric and magnetic fields transfer charge from one Weyl node to the other \cite{Nielsen:1983,Son:2013}. The experimental signature of this phenomenon at low fields---negative magnetoresistance---has been controversial due to the presence of experimental artifacts \cite{Pippard:1989,Reis:2016,Arnold:2016b} and contributions from the trivial hole pocket. We find that using focused-ion-beam lithography (FIB) to reduce current-jetting can suppress or even eliminate the very low field negative magnetoresistance that has been reported for TaAs \cite{Weng:2015,Xu:2015} (see Methods), in line with an earlier study of current-jetting \cite{Reis:2016}. We do observe negative magnetoresistance from 7.5 to 25 tesla which may be related to the chiral anomaly, although not all samples show this to the same degree and this effect may depend sensitively on sample alignment (see Supplementary Methods). Our focus is not on the low-field regime but instead on the QL where the predicted behaviour of the chiral anomaly is qualitatively distinct \cite{Spivak:2016}.

In the QL, where transport along the $z$ direction is determined by the $n=0$ LL of the Weyl fermions alone, the conductivity has the form 
\begin{equation}
\sigma_{zz} = N_W \frac{e^2 v_F}{4 \pi \hbar l_B^2}\tau_{inter}\!\left(B\right),
\label{eq:cond}
\end{equation}
where $N_W$ is the number of Weyl points in the Brillouin zone, $v_F$ is the Fermi velocity, $l_B = \sqrt{\left(h/eB\right)}$ is the magnetic length, and $\tau_{inter}\!\left(B\right)$ is the field-dependent inter-nodal scattering time \cite{Nielsen:1983,Spivak:2016}. As shown below, the W1 Weyl nodes are most likely gapped in this field range (we use the notation of \citet{Arnold:2016} for W1 and W2, opposite that of \citet{Lv:2015b}). We therefore evaluate \autoref{eq:cond} for the eight W2 Weyl nodes, using $v_F$ determined by the quantum oscillations (see Methods). At 27 tesla, where $\sigma_{\mathrm{zz}}$ is maximum, we calculate $\tau_{inter} = 6 \mathrm{~ps}$. This is approximately 10 times longer than the \textit{intra}-nodal scattering time that we calculate from the Dingle factor of the quantum oscillations (see Methods), confirming that the chiral \thh{0} LLs carry current with a reduced scattering rate in the QL due to their physical separation and distinct chiralities. 

As the magnetic field is increased TaAs shows relatively constant resistivity in the QL before 50 tesla (\autoref{fig:rzz}b), requiring that the scattering rate $1/\tau$ increases roughly in proportion to $B$ to cancel the factor of $l_B^2$ in \autoref{eq:cond}. Indeed, $1/\tau \propto B$, and the resulting field-independent conductivity, was predicted for the QL of Weyl semimetals assuming short-range impurity scattering, suggesting that this is the dominant scattering mechanism between the Weyl nodes in TaAs \cite{Spivak:2016}. As pointed out by \citet{Spivak:2016}, short-range impurity scattering in the quantum limit of \textit{conventional} semimetals leads to longitudinal magnetoresistance that increases strongly with magnetic field \cite{Argyres:1956,Murzin:2000}. Thus our observation of relatively constant $\rho_{zz}$ up to 7 times the QL is a unique property of Weyl fermions.

\subsection{Gap formation in the chiral $n=0$ Landau levels}
\begin{figure}
\centering
\includegraphics[width=\textwidth,clip=true,trim=0mm 0mm 0mm 0mm]{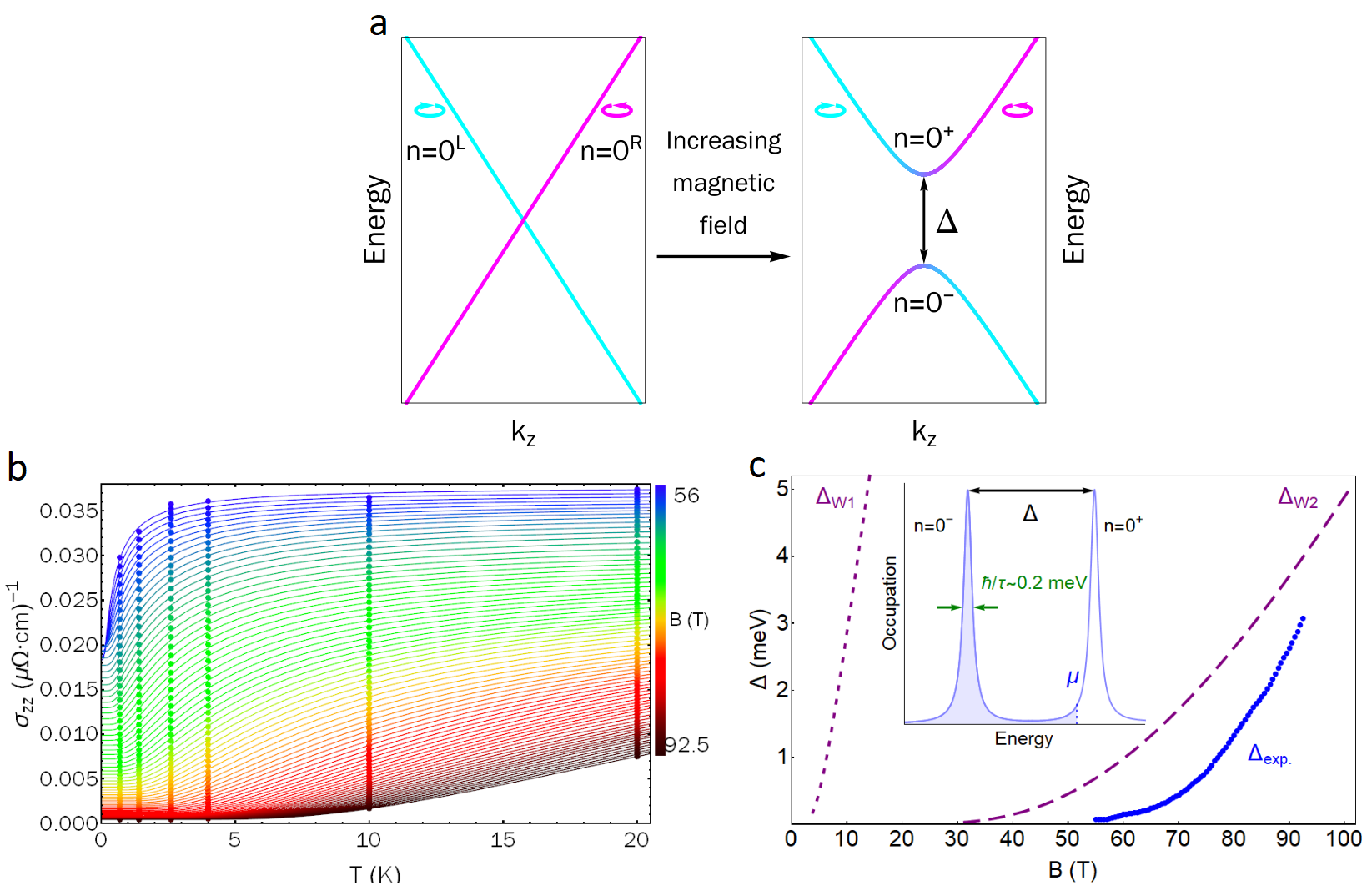}
\caption{\textbf{Mixing of left and right-handed Landau levels in a magnetic field.} The left and right-handed Weyl nodes are separated in the $k_x$-$k_y$ plane in TaAs (\autoref{fig:rzz}a) but are degenerate along $k_z$. At low magnetic fields the left and right-handed $n=0$ LLs can be considered independent (\textbf{a}). As the magnetic field is increased, and the magnetic length approaches the inverse momentum-space separation of the nodes, the two LLs hybridize and a gap $\Delta$ opens that increases with field. We extract this gap by fitting the conductivity to  $\sigma_{zz} = \sigma_0 + \sigma_1 e^{-\frac{\Delta}{k_B T}}$ from 56 to 92.5 tesla (\textbf{b}, points are data and solid lines are fits at different values of the magnetic field), with $\Delta$ show in panel \textbf{c}. We find that the W1 nodes gap at low field due to their close momentum-space proximity \cite{Weng:2015,Lv:2015b,Huang:2015,Ma:2017}, whereas the much larger separation of the W2 nodes means that their gap opens at higher field (\textbf{b}). The $\approx 0.5$~meV offset between our calculated and measured gaps may be due to the $\approx 0.2$~meV LL broading due to finite quasiparticle lifetime (inset). }
\label{fig:mixed}
\end{figure}

Having established the QL transport properties of TaAs at intermediate fields we turn to higher fields where \rh{zz} increases rapidly above 50 tesla (Figure \ref{fig:rzz}b). While the \thh{0} LLs of the Weyl nodes should naively remain metallic in the QL, the two-order-of-magnitude increase in $\rho_{zz}$ suggests the formation of a gap. Weyl nodes are topologically stable to small perturbations \cite{Armitage:2017}, but this can be overcome by strong magnetic fields which mix the left and right-handed $n=0$ LLs (\autoref{fig:mixed}a). If the chemical potential falls within the gap then thermal activation is required to promote charge carriers to the empty LL, resulting in a strong temperature dependence for \rh{zz}. As pointed out previously \cite{Kim:2017,Chan:2017,Zhang:2017c} a gap in the \thh{0} LLs should be observable when the inverse magnetic length $1/l_B = 1/\sqrt{\left(\hbar/eB\right)}$---which controls the momentum-space extent of \thh{0} LL wavefunctions---becomes large compared to the momentum-space separation of the Weyl nodes. This mechanism implies that the left and right handed chiral quasiparticles lose their distinction when their wavefunctions overlap, allowing for increased scattering between them and the formation of a gap. This is the same mechanism proposed for the observed change in sign of the Hall coefficient in TaP \cite{Zhang:2017c}. The significant difference between TaAs and TaP is that one pair of Weyl nodes in TaP remains ungapped up to the highest measured fields, whereas all nodes are eventually gapped in TaAs. The high sensitivity of $\sigma_{\mathrm{zz}}$ to the opening of the gap in TaAs allows us to produce a highly accurate measure of Weyl node separation in the bulk material. 

To quantify this picture we extract the experimental gap by fitting the temperature-dependent conductivity at different values of the magnetic field where the gap opens above 56 tesla (\autoref{fig:mixed}b). The data are well-described by the model $\sigma_{zz} = \sigma_0 + \sigma_1 e^{-\frac{\Delta}{k_B T}}$, where $\sigma_0$ is a field-dependent background, $\sigma_1$ is the conductivity from the \thh{0} LLs, and $\Delta$ is the gap (plotted as $\Delta_{exp}$ in \autoref{fig:mixed}c). This model assumes that the gap dominates the field-dependence of the transport and we keep the scattering rate fixed as function of field for simplicity. To relate this gap to the electronic structure of the Weyl nodes we developed a complete tight binding model of TaAs in a magnetic field, including all 24 Weyl nodes and Zeeman coupling (see Methods for a full description of the tight binding model and LL calculations). Fixing $v_F$ to the experimentally determined value, we find that $\delta k = 0.15 \pi/a$ gives the correct slope of $\Delta$ versus $B$ at high field. The slope of $\Delta$ versus $B$ is a strong function of $\delta k$, and therefore provides a bulk-sensitive measurement of the Weyl node separation. Our extracted value of $\delta k$ also fits the range of W2 node separation estimated by photoemission measurements (between $0.12 \pi/a$ and $0.15 \pi/a$) \cite{Lv:2015,Lv:2015b}. The W1 nodes are estimated to be separated by at most $\delta k = 0.04 \pi/a$, and are therefore gaped at much lower field (\autoref{fig:mixed}c).

It should be noted that perfect electron-hole compensation is required for the chemical potential to lie in the gap at high fields: excess of electrons (holes) will pin the chemical potential in the $n=0^-$ ($n=0^+$) LL to maintain the net carrier concentration (\autoref{fig:mixed}a). Hall effect measurements indicate that compensation in TaAs may not be perfect, with an estimated $2 \times 10^{17}~\mathrm{cm}^{-3}$ to $5 \times 10^{18}~\mathrm{cm}^{-3}$ excess electrons \cite{Zhang:2015,Arnold:2016} that naively should remain metallic at high fields. In reality once the magnetic length is short enough that the left and right nodes overlap scattering will increase greatly and eventually localize any residual carriers. The estimated excess electron density in TaAs is $\approx 10^{12}$ carriers per cm$^2$---within the range of a metal-insulator transition \cite{Popovic:1997,Kravchenko:1994}. This may explain why we observe the gap in $\rho_{zz}$ at slightly higher fields than both our model and the model of \citet{Chan:2017} predicts. Note that we do not observe a similar increase in \rh{xx} above 50 tesla: the in-plane conductivity has an 80 meV (1000 kelvin) cyclotron gap at 50 tesla, and in-plane transport in this regime is dominated by impurity-assisted hopping between adjacent $n=0$ LLs in real space \cite{Murzin:2000}. Thus \rh{xx} should be weakly temperature dependent for $\mathrm{\textbf{B}}||\mathrm{\textbf{c}}$, consistent with what we observe experimentally.

\subsection{Phase transition in the quantum limit}

\begin{figure}
\centering\
\includegraphics[width=\textwidth,clip=true,trim=0mm 0mm 0mm 0mm]{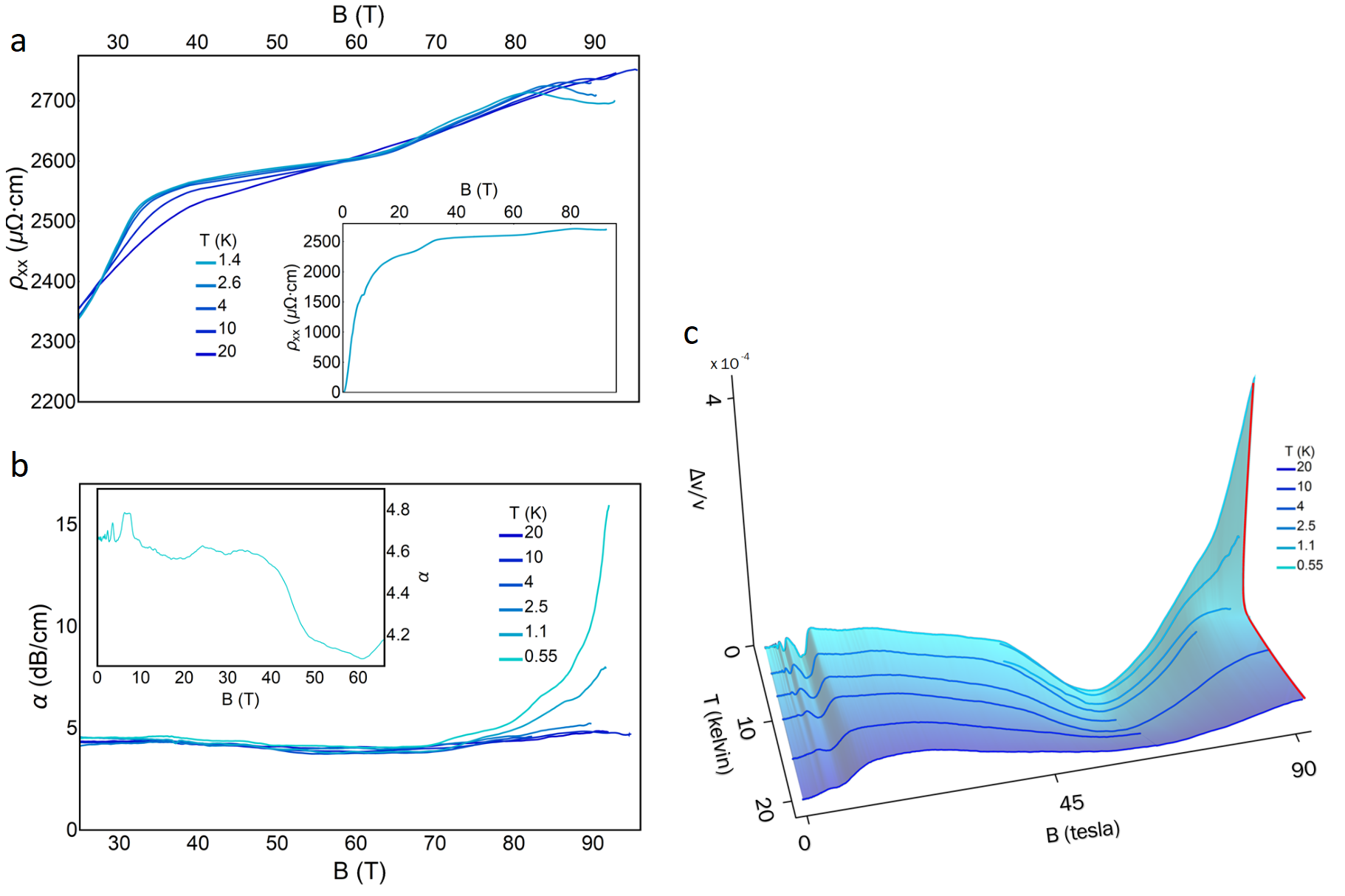}
\caption{\textbf{Phase transition in the quantum limit of TaAs.} \textbf{a}, Resistivity for \bpej, showing a kink near 80 tesla at low temperature. \textbf{b}, The ultrasonic attenuation at 315 MHz for $\mathrm{\textbf{k}}||\mathrm{\textbf{B}}||\mathrm{\textbf{c}}$, where $\mathrm{\textbf{k}}$ is the propagation wavevector of the longitudinal (compressional) sound, and with data from 0 to 65 tesla shown in the inset. \textbf{c}, Change in the speed of sound for the same configuration as the attenuation shown in \textbf{b}. Data from two different experiments---one from 0 to 65 tesla and one from 45 to 95 tesla---are combined in this plot. Above 2.5 K the sound velocity flattens out above 80 tesla and the attenuation is only weakly field dependent. Below 2.5 K, however, both the sound velocity and the ultrasonic attenuation increase rapidly with field. The red line is an interpolation of the data, highlighting the abruptness of the transition as a function of temperature.}
\label{fig:dv}
\end{figure}

At 80 tesla, well into the gapped state, we observe a small decrease in \rh{xx} at the lowest temperatures (\autoref{fig:dv}a)---possibly indicative of a phase transition. To investigate this possibility further we measured the sound velocity ($v_{zz}$) of TaAs using pulse echo ultrasound up to 95 tesla, providing a thermodynamic probe at the highest fields. Below 2.5 K and above 80 tesla we observe an increase in sound velocity accompanied by a strong increase in ultrasonic attenuation (\autoref{fig:dv}b and c). The onset of these ultrasonic features coincides with the sharp decrease in \rh{xx}, and we take all three phenomena as evidence for a field-induced phase transition. Our numerical solutions of the tight binding model predict no further transitions as a function of magnetic field once the gap opens along $k_z$, suggesting that this transition at 80 tesla is driven by interactions. 

There is precedent for an interaction-driven phase transition in the quantum limit of a semimetal. Graphite, whose quantum limit is 7.5 tesla, undergoes a rich series of transitions above 30 tesla. These transitions are generally ascribed to excitonic or Peirels-type density-wave formation along the magnetic field direction \cite{Tanuma:1981,Fauque:2013,Zhu:2017}. This mechanism seems unlikely to occur in TaAs given that a gap has already opened along $k_z$ at 50 tesla. In the limit where the ultrasonic wavelength is much longer than the quasiparticle mean free path, attenuation is proportional to conductivity \cite{Pippard:1955}. The fact that we observe attenuation to increase while $\sigma_{zz}$ \textit{decreases} suggests that something other than conventional electronic attenuation is occurring above 80 T in TaAs. The enhanced attenuation could come from the interaction between ultrasound and order-parameter fluctuations, as it does in graphite where it peaks near each field-induced phase transition \cite{LeBoeuf:2017}. The unbounded growth in attenuation we observe suggests that either we have not yet reached the attenuation peak, or that a different attenuation mechanism is at work.

To speculate on the microscopic origin of the high-field phase it is helpful to consider which degree of freedom remains after the gap opens along $k_z$. The in-plane cyclotron orbits of the quasiparticles shrink in size with increasing magnetic field, resulting in a large real-space degeneracy of the $n=0$ LLs (or equivalently momentum-space overlap of the quasiparticle wavefunctions). This increases the Coulomb repulsion between quasiparticles as $E_C \propto e^2/l_B$, ultimately leading to Wigner crystallization. Before this transition occurs, however, there are intermediate `mixed' or `microemulsion' phases with spatially phase-separated regions of Fermi liquid and bubbles of crystallized electrons \cite{Spivak:2006}. Disorder pins these structures in real-space, resulting in a mesoscopically inhomogeneous system. This texture, with a length-scale comparable to the ultrasonic wavelength, is known to scatter ultrasound. Clearly this phase needs to be explored further before it can be identified: the frequency dependence of the attenuation would yield the length scale of the phase separation \cite{Paalanen:1992} or the frequency-dependence of the order parameter fluctuations \cite{Shekhter:2013}; the collective sliding motion of crystallized electrons can be revealed by nonlinear current-voltage measurements \cite{Field:1986}.

\section{Discussion}

The ideal Weyl semimetal contains only one pair of Weyl nodes, with large momentum space separation to make them robust against perturbations. Reality is more interesting: the W2 Weyl nodes in TaAs have a large separation compared to other materials \cite{Weng:2015}, but there are 12 pairs of nodes in total plus the trivial hole pockets. Using magnetic fields to drive Weyl semimetals into their QL provides a clear path forward for studying `pure' Weyl physics, taking advantage of the fact that the \thh{0} LLs are field-independent. This greatly broadens the number of potential systems where Weyl fermions can be accessed: as long as the magnetic field is strong enough the chemical potential will always move to the chiral \thh{0} Weyl LLs in the quantum limit, even if the zero-field Fermi surface encompasses both nodes. Given that magnetic fields also increase the Coulomb interaction between quasiparticles (which can be particularly weak in semimetals due to the high Fermi velocities), the QL is also a promising regime for finding new states of matter formed from Weyl fermions.

\section{Methods}

\subsection{Sample preparation}
\label{se:samp}
Negative magnetoresistance has been reported for a large number of candidate Dirac and Weyl semimetals \cite{Armitage:2017}. The observation of decreasing resistance with increasing magnetic field for $\mathrm{\textbf{J}}||\mathrm{\textbf{B}}$ has been taken as evidence of the chiral anomaly. Some of these measurements have been disputed because of extraneous effects that can also lead to an apparent negative magnetoresistance. In particular for TaAs, the highly anisotropic nature of the magnetoresistance means that current injected into the sample through point-like contacts does not necessarily move uniformly through the sample. This effect has been known for quite some time (see page 43 of Pippard \cite{Pippard:1989}), and recent numerical simulations explored this in the context of Weyl semimetals by \citet{Reis:2016}. The result is that the potential drop across a pair of voltage contacts may not reflect the total current injected into the sample, depending on where the contacts are placed. This can be overcome by avoiding point-like contacts and by preparing samples with high aspect ratios with current injection taking place far from the voltage measurement. Even a standard precaution such as contacting the entire ends of a sample to inject current homogeneously does not necessarily produce consistent results since microscopic contact to the sample can still be point like or at the very least inhomogeneous. 

To illustrate this point for TaAs we prepared the sample shown in Supplementary Figure 1. The current path is highlighted in purple; the `good' voltage contacts in red; the `bad' voltage contacts in green. The corresponding resistances, or rather the voltage drop across the contacts divided by the nominal current, are shown in the two right hand panels for $\mathrm{\textbf{B}}||\mathrm{\textbf{J}}||\mathrm{\textbf{c}}$: the top panel corresponds to the green-shaded contacts offset from the current path; the bottom panel corresponds to the red-shaded contacts that are closer to the current path. It is immediately clear that a sizable ``negative magnetoresistance'' is visible for the green contacts at all temperatures, similar in character to that reported as evidence for a chiral anomaly in TaAs by Zhang \textit{et al.} \cite{Zhang:2016} and Huang \textit{et al.} \cite{Huang:2015}. In our exaggerated geometry the voltage drops to near-zero at high fields: clearly an unphysical result. The apparent increase in this effect at higher temperatures is due to the increase in resistance and decrease in resistive anisotropy at higher temperatures: it takes more magnetic field to isolate the current from the green voltage contacts at higher temperatures because the resistive anisotropy is lower. The red contacts, on the other hand, show positive magnetoresistance that saturates at approximately 1 tesla---this is expected as $\omega_c \tau$ reaches 1 near this field value. We used an even more restricted current path geometry for the data presented in the main text, and a second sample with similar current path characteristics shows similar resistivity in Supplementary Figure 3.

\rh{zz} shown in Figure 1 of the main text show signs of saturation at the highest resistance values. This may be due to a parallel conduction channel on the surface of the sample. There are two possible sources of this conduction channel: the surface states which are known to be present on this material and which have been observed via ARPES \cite{Lv:2015}; and/or an amorphous arsenic-depleted layer induced by the FIB \cite{Bachmann:2017}.

As can be clearly seen in the inset of Figure 1 in the main text, the FIB prepared samples undergo a superconducting transition at low temperatures and low magnetic field. This is also a surface effect: it was recently shown that after FIB microstructuring the related material NbAs develops an amorphous arsenic-depleted surface layer, as arsenic is preferentially removed during the FIB process, leaving behind a niobium (or in our case tantalum) rich layer that is superconducting \cite{Bachmann:2017}. This superconductivity is fully suppressed above 2 tesla and 1.5 kelvin, and thus does not affect the majority of the data presented.

While the presence of a conducting surface layer changes the shapes of our resistivity curves at high fields, it does not alter any of the features from which we draw our main conclusions: the positions of the quantum limits for the various pockets, the onset of the resitivity increase in \rh{zz} near 50 tesla, the temperature dependence of the increase in \rh{zz}, the lack of a strong increase near 50 tesla in \rh{xx}, and the sharp downturn in \rh{xx} near 80 T below 4 K. The consistency of the ultrasound with the features in the transport provides further confirmation of the intrinsic nature of these features, as there was no FIB lithography performed on the ultrasound sample.

\subsection{Fits to the conductivity}
\label{se:gap}
We fit the conductivity $\sigma_{zz} (= 1/{\rho_{zz}}$) at each field slice as a function of temperature to a simple mode: a background $\sigma_b$ plus an activated component
\begin{equation}
\sigma_{zz} = \sigma_b + \sigma_0 e^{-\alpha/T},
\label{eq:gap}
\end{equation}
where $\alpha$ is the gap in units of kelvin. These fits plus the field-dependent fit parameters are shown in Supplementary Figure 2. 

In order to fit the gap down to the lowest possible field we normalized the conductivity to its value at 47 tesla. Supplementary Figure 2 demonstrates that while this procedure changes $\sigma_b$ and $\sigma_0$ it leaves $\Delta$ relatively unchanged. This is because the exponential factor dominates the fit, producing variability in $\sigma_b$ and $\sigma_0$ that depend on how (which field value, for example) the normalization is carried out but leaving $\Delta$ procedure-independent. While $\sigma_0$ is a physical parameter representing background conductivity in TaAs, the extracted values of $\sigma_b$ and $\sigma_0$ should not be taken as intrinsic. 

\subsection{Quantum oscillation analysis}
\label{se:oscillations}

According to the detailed quantum oscillation studies performed by \citet{Arnold:2016}, TaAs contains three distinct types of Fermi surface: electron-like W1 Weyl surfaces, electron-like W2 Weyl surfaces, and hole-like trivial surfaces. We do not attempt to re-create their entire analysis here, but instead focus on the oscillations we observe in the \bpaj transport geometry. A complete treatment of the oscillation frequencies requires a full angle dependence which we have not measured: the purpose of the following analysis is to obtain a reasonable estimate for the quasiparticle lifetime $\tau$ for the Weyl electrons and to provide further evidence that the $\approx 18$~T oscillation frequency is indeed coming from the same hole pocket identified by \citet{Arnold:2016}. 

The full $\hat{c}$-axis conductivity, including both background and oscillatory contributions, can be written at fixed angle as
\begin{equation}
\sigma_{zz} = \sum_i{\sigma^i_0\!\left(B\right)\left(1+\sum_p{A^{i,p} R^{i,p}_T R^{i,p}_D \cos\!\left(2\pi p \frac{F_i}{B}+\phi_i\right)}\right)},
\label{eq:osc}
\end{equation}
where the first sum is over the $i$ distinct Fermi surfaces, the sum $p$ is over the harmonics, $\sigma^i_0\!\left(B\right)$ are the field-dependent background conductivities, $A^{i,p}$ are amplitude factors, $F_i$ are the oscillation frequencies (proportional to the Fermi surface cross sectional area), and $\phi_i$ are phase factors \cite{Shoenberg:1984}. The temperature factor is given by 
\begin{equation}
R^{i,p}_T = \frac{\frac{2\pi^2 p k_B T}{\hbar \omega_c^i}}{\sinh\!\left(\frac{2\pi^2 p k_B T}{\hbar \omega_c^i}\right)},
\label{eq:temper}
\end{equation}
where $\omega_c^i = e B/m_i^{\star}$ is the cyclotron frequency for effective mass $m_i^{\star}$. The Dingle factor is
\begin{equation}
R^{i,p}_D = e^{-\frac{\pi p}{\omega_c^i \tau_i}},
\label{eq:dingle}
\end{equation}
where $\tau_i$ is the quasiparticle lifetime. 

It is important to note that the background conductivities in \autoref{eq:osc} are multiplicative with the oscillation amplitudes, not additive. Thus if a piece of Fermi surface does not contribute to $\sigma_{zz}$ then its oscillations will not be visible in \rh{zz}, even if the quasiparticle lifetime for that surface is long. The W1 pockets are roughly elongated ellipses (see Figure 3C of \citet{Arnold:2016}) with an aspect ratio of approximately 10:1 with the long axis oriented along the $\hat{c}$ direction. With an in-plane Fermi velocity of $3\times10^5$~m/s these surfaces dominate the \rh{xx} transport and produce strong Shubnikov-de Haas oscillations with $\mathrm{\textbf{B}} || \mathrm{\textbf{c}}$. For the \rh{zz} transport, however, we expect their Fermi velocity along $\hat{c}$ to be reduced by the aspect ratio (a factor of 10, so $v_Z \approx 3\times 10^4$~m/s), and thus contribute very little to the \rh{zz} transport. The W2 Weyl surfaces, on the other hand, are roughly isotropic in aspect ratio and have $v_z \approx 1.6\times 10^5$~m/s; the hole pocket has $v_z \approx 7\times 10^5$~m/s. As conductivity is proportional to Fermi velocity squared, \rh{zz} should be dominated by the W2 electron and trivial hole pockets. Note that the oscillations from a particular pocket can still be small due to other factors: for example, the hole pockets for $B||\hat{c}$ have a very unfavourable curvature factor that reduces their oscillation amplitude considerably (oscillation amplitudes scale as $1/\sqrt{\partial^2 A / \partial \kappa^2}$, where $A$ is the extremal area of the Fermi surface perpendicular to the magnetic field and $\kappa$ is the momentum direction parallel to the magnetic field) \cite{Shoenberg:1984}. 

We forgo the usual Fourier analysis of our data which can be difficult when there are only a few oscillations \cite{Ramshaw:2011}. We instead perform a full three-dimensional fit to \autoref{eq:osc} for our data at temperatures where superconductivity is suppressed ($T=2.6,4,10$ and $20$~K). We can account for most features in the data fitting with $F_{W2} = 6.5$~T, $F_{H} = 17.2$~T, $m_{W2}^{\star}=0.1$~$m_e$, $m_{H}^{\star}=0.2$~$m_e$, $\tau_{W2} = 0.6$~ps, and $\tau_{H} = 0.4$~ps Supplementary Figure 5. Values of $F_{W2} = 5.5$~T and $m_{W2}^{\star}=0.105$~$m_e$ were obtained by \citet{Arnold:2016} for the W2 Weyl pocket measuring magnetic torque with field along the $\hat{a}$ direction: given the relatively isotropic nature of this pocket we believe that we are observing the same piece of Fermi surface. Values of $F_{H} = 18.8$~T, and $\tau_{H} = 0.11$~ps were obtained for the hole pocket by \citet{Arnold:2016}, which again are in reasonable agreement with our own. Note that \citet{Arnold:2016} were unable to obtain a mass for the hole pocket for $\mathrm{\textbf{B}}||\mathrm{\textbf{c}}$ but predicted a mass of $m_H^{\star} = 0.17$~$m_e$---very close to our experimentally determined $m_{H}^{\star}=0.2$~$m_e$ for this pocket.

\subsection{Pulse echo ultrasound}
\label{se:pulse}

Pulse echo ultrasound measures the speed of sound $v$ transmitted through a sample for a specified sound propagation direction and polarization. Elastic moduli can then be computed via the relation $v = \sqrt{\frac{c}{\rho}}$, although care must be taken when choosing the propagation and polarization directions if the relationship between $v$ and any particular $c_{ij}$ is to be kept simple (see Brugger \textit{et al.} \cite{Brugger:1965}). The measurement typically involves generating a short burst of ultrasound (of order a few hundred nanoseconds) via a piezoelectric transducer, and then recording the echoes with the same transducer as the sound travels back and forth between the faces of the sample parallel to the transducer face. The phase shift between each successive echo is related to the travel time (and hence sound velocity) of the sound pulse, and the amplitude decay with successive echoes is related to the ultrasonic attenuation. 

Our implementation of pulse echo ultrasound was similar to Suslov \textit{et al.} \cite{Suslov:2006} with one notable exception: we forwent the mixing stage and directly recorded the full ultrasonic waveform with a 6 GS/s digitizer card (GaGe EON Express). This has two major advantages: it eliminates a large fraction of the electronics; and it allows the use of digital filtering and lock-in techniques that can be optimized and re-processed after the experiment is completed.

We generated 315 MHz longitudinal ultrasound with a \degr{36} Y-cut LiNbO$_3$ transducer affixed to a $(0,0,1)$ face of a TaAs sample (Supplementary Figure 6). As both the direction of propagation and the polarization are along $\hat{c}$, we probe purely the $c_{33}$ elastic modulus in this tetragonal crystal \cite{Brugger:1965}. The full field dependence at selected temperatures between 0.55 and 20 kelvin is shown in Supplementary Figure 6. The velocity $v$ is calculated from the phase shift $\phi$ between successive echoes as 
\begin{equation}
\Delta v = -2\pi f l \frac{\Delta \phi}{\phi^2},
\label{eq:vel}
\end{equation}
where $f$ is the ultrasonic frequency and $l$ is twice the sample length. The absolute phase $\phi$ can be obtained from the absolute spacing between successive echoes (a less accurate measure of the sound velocity, but the only way to get the absolute value) via $v = 2\pi f l/\phi$. All plotted velocity shifts are normalized to the phase shift at $B=0$: the absolute shift as a function of temperature was not obtained.

\subsection{Tight-binding model for TaAs}
\label{se:tight}

The primitive unit cell of TaAs consists of four sublattices, two Ta atoms and two As atoms, per primitive unit cell. The lattice structure has two mirror symmetries $M_x$ and $M_y$, TRS, and is invariant under $C_4$ rotation followed by a translation along the $\hat z$ axis. We consider a three-dimensional tight-binding model representing the TaAs lattice structure. Specifically, we consider one spinful orbital per lattice site, giving rise to eight bands in total. 

Our tight binding model:
\begin{eqnarray}
H &=& t\sum_{\langle ij \rangle,s}c^\dagger_{is} c_{js}  + \sum_i \Delta_i c^\dagger_i c_i
\nonumber\\&+&i\lambda \sum_{\langle \langle ik \rangle \rangle,s,s'}c^\dagger_{is} c_{ks'}\sum_{j}[\hat {d}_{ijk}\cdot\vec\sigma_{ss'}]
\label{eq:mintbm}
\end{eqnarray}
where $t$ is the nearest neighbor hopping; $\Delta$ is a staggered potential: $\Delta_i = \pm \Delta$ depending on whether the lattice site contains Ta or As; $\lambda$ is the amplitude of the spin orbit interaction between the next nearest neighbors; $s=\uparrow,\downarrow$ is the spin; and $\sigma$'s are the corresponding Pauli matrices. $\hat d_{ijk}$ is the unit vector in the direction of $\vec d_{ij}\times \vec d_{jk}$, where $j$ is an intermediate site between $i$ and $k$. This Hamiltonian preserves time reversal symmetry. 

\subsubsection{Weyl semimetal symmetries and the number of Weyl points}

TaAs has 24 Weyl points, of which 8 are in the $k_z=0$ plane (W1) and the rest 16 (W2) are at finite $k_z$ \cite{Weng:2015}. The 8-fold (16-fold) multiplicity of the W1 (W2) Weyl points are guaranteed by model symmetries (lattice and TRS). More generally, the total number of Weyl points has to be $8n_1+16n_2$, where $n_1$ and $n_2$ are the numbers of sets of W1 and W2 Weyl points, respectively. For example, we will show the minimal tight-binding model can give 8 Weyl points, while adding extra terms and anisotropy reproduces the 24 Weyl points. 

A finite $\Delta$ in Eq. \ref{eq:mintbm} explicitly breaks inversion symmetry, thus is vital to the emergence of Weyl semimetals. On the other hand, in the limit of $\Delta \rightarrow \infty$ the system has to be a trivial insulator. In the presence of TRS, a Weyl semimetal is guaranteed to exist between a topological insulator and a trivial insulator\cite{Turner:2012}. During the evolution, pairs of Weyl points are generated, transported across the momentum space, and re-annihilate in different pairs, giving rise to a change of the $\mathbf Z_2$ topological index. We note that the W2 Weyl points always change the topological index trivially by 2, and it is up to the W1 Weyl points to change the $\mathbf Z_2$ topological index by a nontrivial 1. This suggests that while the W1 Weyl points are essential and guaranteed to occur somewhere during the interpolation, the W2 Weyl points are less fundamental and involve more fine-tuning. 

\subsubsection{Results}

First, we show that the Hamiltonian in Eq. \ref{eq:mintbm} can give a Weyl semimetal in certain parameter regions. For model parameters $t=1.0$, $\Delta=0.7$, and $\lambda=0.1$, the dispersion around $\vec k=(0.7, 2.56, 0)$ is shown in Supplementary Figure 7. The band touching at a single Weyl point and the nearby linear dispersion are clearly established. The locations of all the eight Weyl points within the momentum space are illustrated in Supplementary Figure 8. Being at $k_z=0$, these Weyl points belong to the W1 group. Interestingly, we observe the shift and annihilation of the Weyl nodes as we change the amplitude $\Delta$ of inversion symmetry breaking. We also note that the dispersion between the pair of adjacent Weyl nodes resembles that observed in band-structure calculations (Supplementary Figure 9)\cite{Ma:2017}. 

A comparison can be made between the electronic structure of the minimum tight-binding model Supplementary Figure 10 and that of the actual TaAs crystal from band-structure calculations (see for example figure 4c from Ref. \cite{Huang:2015}). The minimum model in Eq. \ref{eq:mintbm} is already capable of qualitatively resembling the ab-initio band structures near the chemical potential. 

\subsubsection{Consequences of a magnetic field}

We now consider the fate of the Weyl physics in the presence of an external magnetic field. For simplicity, we first focus on the situation when the magnetic field is along the c axis: $\vec B = B_z \hat{z}$. Since $k_z$ remains a good quantum number, together with the quantization within the $k_x-k_y$ plane, gives rise to Landau bands. For an individual Weyl node, there exists a chiral Landau band, whose up-moving or down-moving is associated with the chirality of the original Weyl node. However, as discussed in other theoretical studies, this physics becomes challenged at large magnetic fields\cite{Kim:2017,Chan:2017}.  

\subsubsection{Anti-crossing of the chiral Landau levels}

Semi-classically, magnetic fields introduce finite resolution in $k_x -k_y$ momentum space, inversely proportional to the magnetic length. Therefore, at large enough $B$, the pair of separated Weyl nodes with opposite chirality becomes indistinguishable and magnetic breakdown makes tunneling between them non-negligible. This was addressed schematically in Ref. \cite{Kim:2017,Chan:2017}. Here, we will study from the perspective of our tight-binding model in Eq. \ref{eq:mintbm}. We find an emergent gap whose amplitude and onset field are roughly consistent with those observed in experiments for Weyl node separation $\delta k \sim 0.15 \pi/a$, which is at least a factor of 2 larger than DFT results\cite{Ma:2017,Huang:2015} yet can be attributed to a relatively small modification to the overall band structure. Since the separation between the pair of adjacent Weyl fermions are small, they are very sensitive to small shifts in bandstructure which may be beyond the resolution capability of the DFT calculations. We also note that the Weyl nodes separation $\delta k \sim 0.15 \pi/a$ is consistent with the position of the Weyl nodes in the surface Brillouin zone observed in ARPES experiments\cite{Lv:2015b}. As most studies have found W2 Weyl fermions to have larger separations than the W1 Weyl fermions, the former should be more resilient and dominates the fate of the Weyl physics in large magnetic fields. Although the W2 Weyl nodes are away from the $k_z=0$ plane, it only causes a constant shift to the zero of the $k_z$ good quantum number and does not change the following discussion. The $k$-space distance between different pairs of Weyl nodes is much greater than the intra-pair separation and quantum tunneling due to magnetic breakdown between different pairs can be neglected. 

In this subsection, we neglect the Zeeman effect, and the vector potential of the magnetic field is chosen as $\vec A = (0, Bx, 0)$. We also set Fermi velocities in the $v_F^x\sim v_F^y\sim 291,000 m/s = 1.92 eV\cdot \AA$ observed via quantum oscillation measurements. These, together with the Weyl node separation $\delta k \sim 0.15 \pi/a$ and $a=3.437\AA$ in TaAs allows us to set proper units for our numerical results.

A typical Landau band dispersion is illustrated in Supplementary Figure 11. Each band is four-fold degenerate, suggesting the tunneling between different pairs of Weyl nodes are still negligible even at a large magnetic field $B_z=145T$. On the other hand, a gap is clearly observable between the $n=0$ chiral Landau bands. 

An important question is how does the size of this anti-crossing gap depends on the applied magnetic field. Intuitively, the gap should be exponentially suppressed at smaller field $l_B \delta_k \gg 1$, and scale linearly at larger magnetic field. Such behavior is clearly observed in Supplementary Figure 12 and Supplementary Figure 13, where the units of the data are interpreted for TaAs W2 Weyl nodes with separation $\delta k=0.15\pi/a$ and $\delta k=0.165\pi/a$, respectively. 

\subsubsection{Impact of the Zeeman effect}

Now we consider another potentially important ingredient---the Zeeman effect---particularly important because semi-metals and semi-conductors often give rise to large $g$-factors. For example, the $g$-factor of TaP has been estimated to be between 2 and 2.9 for one set of orbits and 5.5 and 6.7 for a second set \cite{Hu:2016}. To incorporate this we add an additional contribution to the Hamiltonian: 
\begin{eqnarray}
H_{\mbox{Zeeman}}=-\mu_B g S^z 
\label{eq:Zeeman}
\end{eqnarray}

The results with and without the Zeeman effect are compared in Supplementary Figure 14. The results suggest that though the impact of the Zeeman effect is relatively minor (Supplementary Figure 14 upper panel), the quantitative change to the anti-crossing feature is visible especially for larger Weyl nodes separation $\delta k$ that allows weaker gaps with a larger magnetic field threshold, see Supplementary Figure 14 lower panel.

The influence of the Zeeman effect can be understood via its modifications to the band structure in the absence of the vector potentials, see Supplementary Figure 15. In the presence of Zeeman energy, the Weyl nodes moves further apart, making quantum tunneling between the pair of Weyl nodes more difficult thus suppressing the anti-crossing gap size. 

\section*{Data Availability}
The data that support the findings of this study are available from the corresponding author upon reasonable request.

\section*{Acknowledgments}
This work was performed at the National High Magnetic Field Laboratory, which is supported by National Science Foundation Cooperative Agreement No. DMR-1157490 and the State of Florida. E.D.B, F.R., and R.D.M. acknowledge funding from the LANL LDRD DR20160085 `Topology and Strong Correlations'. B.J.R. acknowledges funding from LANL LDRD 20160616ECR `New States of Matter in Weyl Semimetals'. M.K.C. acknowledges funding from the U.S. Department of Energy Office of Basic Energy Sciences “Science at 100 T” program. P.J.M. is supported by the Max-Planck-Society and the European Research Council (ERC) under the European Union's Horizon 2020 research and innovation programme (grant agreement No. 715730). M.D.B. acknowledges studentship funding from EPSRC under grant no. EP/I007002/1.
 YZ was supported in part by NSF DMR-1308089 and EAK was supported in part by National Science Foundation (Platform for the Accelerated Realization, Analysis, and Discovery of Interface Materials (PARADIM)) under Cooperative Agreement No. DMR-1539918. 
\section*{Author Contributions}
B.J.R. and R.D.M. conceived of the experiment, E.D.B., N.J.G., and F.R. grew the samples, B.J.R. and R.D.M. performed the focused ion beam lithography, B.J.R., J.B.B., A.M., F.B., and R.D.M. designed and built the experiment, B.J.R., K.A.M., P.J.W.M., M.B., M.K.C., J.B.B., and R.D.M. performed the experiment, B.J.R. and K.A.M. analyzed the data, Y.Z. and E.A.K. developed the tight-binding model and provided theoretical guidance, and B.J.R. wrote the manuscript with input from all coauthors. 

\section*{Competing Interests}
The authors declare no competing interests.

\newpage

\begin{thebibliography}{56}
\providecommand{\natexlab}[1]{#1}
\providecommand{\url}[1]{\texttt{#1}}
\expandafter\ifx\csname urlstyle\endcsname\relax
  \providecommand{\doi}[1]{doi: #1}\else
  \providecommand{\doi}{doi: \begingroup \urlstyle{rm}\Url}\fi

\bibitem[Onnes(1911)]{Onnes:1911}
H~Kamerlingh Onnes.
\newblock {The superconductivity of mercury}.
\newblock \emph{Comm. Phys. Lab. Univ. Leiden}, 122:\penalty0 124, 1911.

\bibitem[Bednorz and M{\"u}ller(1986)]{Bednorz:1986}
J~George Bednorz and K~Alex M{\"u}ller.
\newblock {Possible high-Tc superconductivity in the Ba-La-Cu-O system}.
\newblock \emph{Zeitschrift f{\"u}r Physik B Condensed Matter}, 64\penalty0
  (2):\penalty0 189--193, 1986.

\bibitem[Kivelson et~al.(1998)Kivelson, Fradkin, Emery, et~al.]{Kivelson:1998}
SA~Kivelson, E~Fradkin, VJ~Emery, et~al.
\newblock {Electronic liquid-crystal phases of a doped Mott insulator}.
\newblock \emph{Nature}, 393\penalty0 (6685):\penalty0 550--553, 1998.

\bibitem[Tsui et~al.(1982)Tsui, Stormer, and Gossard]{Tsui:1982}
D.~C. Tsui, H.~L. Stormer, and A.~C. Gossard.
\newblock {Two-Dimensional Magnetotransport in the Extreme Quantum Limit}.
\newblock \emph{Phys. Rev. Lett.}, 48:\penalty0 1559--1562, May 1982.
\newblock \doi{10.1103/PhysRevLett.48.1559}.
\newblock URL \url{http://link.aps.org/doi/10.1103/PhysRevLett.48.1559}.

\bibitem[Mott(1968)]{Mott:1968}
NF~Mott.
\newblock {Metal-insulator transition}.
\newblock \emph{Reviews of Modern Physics}, 40\penalty0 (4):\penalty0 677,
  1968.

\bibitem[Wan et~al.(2011)Wan, Turner, Vishwanath, and Savrasov]{Wan:2011}
Xiangang Wan, Ari~M. Turner, Ashvin Vishwanath, and Sergey~Y. Savrasov.
\newblock {Topological semimetal and Fermi-arc surface states in the electronic
  structure of pyrochlore iridates}.
\newblock \emph{Phys. Rev. B}, 83:\penalty0 205101, May 2011.
\newblock \doi{10.1103/PhysRevB.83.205101}.
\newblock URL \url{https://link.aps.org/doi/10.1103/PhysRevB.83.205101}.

\bibitem[Weng et~al.(2015)Weng, Fang, Fang, Bernevig, and Dai]{Weng:2015}
Hongming Weng, Chen Fang, Zhong Fang, B.~Andrei Bernevig, and Xi~Dai.
\newblock {Weyl Semimetal Phase in Noncentrosymmetric Transition-Metal
  Monophosphides}.
\newblock \emph{Phys. Rev. X}, 5:\penalty0 011029, Mar 2015.
\newblock \doi{10.1103/PhysRevX.5.011029}.
\newblock URL \url{https://link.aps.org/doi/10.1103/PhysRevX.5.011029}.

\bibitem[Xu et~al.(2015)Xu, Belopolski, Alidoust, Neupane, Bian, Zhang, Sankar,
  Chang, Yuan, Lee, Huang, Zheng, Ma, Sanchez, Wang, Bansil, Chou, Shibayev,
  Lin, Jia, and Hasan]{Xu:2015}
Su-Yang Xu, Ilya Belopolski, Nasser Alidoust, Madhab Neupane, Guang Bian,
  Chenglong Zhang, Raman Sankar, Guoqing Chang, Zhujun Yuan, Chi-Cheng Lee,
  Shin-Ming Huang, Hao Zheng, Jie Ma, Daniel~S. Sanchez, BaoKai Wang, Arun
  Bansil, Fangcheng Chou, Pavel~P. Shibayev, Hsin Lin, Shuang Jia, and M.~Zahid
  Hasan.
\newblock {Discovery of a Weyl fermion semimetal and topological Fermi arcs}.
\newblock \emph{Science}, 349\penalty0 (6248):\penalty0 613--617, 2015.

\bibitem[Lv et~al.(2015{\natexlab{a}})Lv, Weng, Fu, Wang, Miao, Ma, Richard,
  Huang, Zhao, Chen, Fang, Dai, Qian, and Ding]{Lv:2015}
B.~Q. Lv, H.~M. Weng, B.~B. Fu, X.~P. Wang, H.~Miao, J.~Ma, P.~Richard, X.~C.
  Huang, L.~X. Zhao, G.~F. Chen, Z.~Fang, X.~Dai, T.~Qian, and H.~Ding.
\newblock {Experimental Discovery of Weyl Semimetal TaAs}.
\newblock \emph{Phys. Rev. X}, 5:\penalty0 031013, Jul 2015{\natexlab{a}}.
\newblock \doi{10.1103/PhysRevX.5.031013}.
\newblock URL \url{http://link.aps.org/doi/10.1103/PhysRevX.5.031013}.

\bibitem[Nielsen and Ninomiya(1983)]{Nielsen:1983}
Holger~Bech Nielsen and Masao Ninomiya.
\newblock {The Adler-Bell-Jackiw anomaly and Weyl fermions in a crystal}.
\newblock \emph{Physics Letters B}, 130\penalty0 (6):\penalty0 389--396, 1983.

\bibitem[Wei et~al.(2012)Wei, Chao, and Aji]{Wei:2012}
Huazhou Wei, Sung-Po Chao, and Vivek Aji.
\newblock {Excitonic Phases from Weyl Semimetals}.
\newblock \emph{Phys. Rev. Lett.}, 109:\penalty0 196403, Nov 2012.
\newblock \doi{10.1103/PhysRevLett.109.196403}.
\newblock URL \url{https://link.aps.org/doi/10.1103/PhysRevLett.109.196403}.

\bibitem[Zhang and Shindou(2017)]{Zhang:2017}
Xiao-Tian Zhang and Ryuichi Shindou.
\newblock {Transport properties of density wave phases in three-dimensional
  metals and semimetals under high magnetic field}.
\newblock \emph{Physical Review B}, 95\penalty0 (20):\penalty0 205108, 2017.

\bibitem[Cho et~al.(2012)Cho, Bardarson, Lu, and Moore]{Cho:2012}
Gil~Young Cho, Jens~H. Bardarson, Yuan-Ming Lu, and Joel~E. Moore.
\newblock {Superconductivity of doped Weyl semimetals: Finite-momentum pairing
  and electronic analog of the ${}^{3}$He-$A$ phase}.
\newblock \emph{Phys. Rev. B}, 86:\penalty0 214514, Dec 2012.
\newblock \doi{10.1103/PhysRevB.86.214514}.
\newblock URL \url{https://link.aps.org/doi/10.1103/PhysRevB.86.214514}.

\bibitem[Meng and Balents(2012)]{Meng:2012}
Tobias Meng and Leon Balents.
\newblock {Weyl superconductors}.
\newblock \emph{Phys. Rev. B}, 86:\penalty0 054504, Aug 2012.
\newblock \doi{10.1103/PhysRevB.86.054504}.
\newblock URL \url{https://link.aps.org/doi/10.1103/PhysRevB.86.054504}.

\bibitem[Zhang and Nagaosa(2017)]{Zhang:2017b}
Xiao-Xiao Zhang and Naoto Nagaosa.
\newblock {Tomonaga-Luttinger liquid and localization in Weyl semimetals}.
\newblock \emph{Phys. Rev. B}, 95:\penalty0 205143, May 2017.
\newblock \doi{10.1103/PhysRevB.95.205143}.
\newblock URL \url{https://link.aps.org/doi/10.1103/PhysRevB.95.205143}.

\bibitem[Armitage et~al.(2018)Armitage, Mele, and Vishwanath]{Armitage:2017}
N.~P. Armitage, E.~J. Mele, and Ashvin Vishwanath.
\newblock {Weyl and Dirac semimetals in three-dimensional solids}.
\newblock \emph{Rev. Mod. Phys.}, 90:\penalty0 015001, Jan 2018.
\newblock \doi{10.1103/RevModPhys.90.015001}.
\newblock URL \url{https://link.aps.org/doi/10.1103/RevModPhys.90.015001}.

\bibitem[Adler(1969)]{Adler:1969}
Stephen~L. Adler.
\newblock {Axial-Vector Vertex in Spinor Electrodynamics}.
\newblock \emph{Phys. Rev.}, 177:\penalty0 2426--2438, Jan 1969.
\newblock \doi{10.1103/PhysRev.177.2426}.
\newblock URL \url{https://link.aps.org/doi/10.1103/PhysRev.177.2426}.

\bibitem[Bell and Jackiw(1969)]{Bell:1969}
John~S Bell and Roman Jackiw.
\newblock {A PCAC puzzle: $\pi$0→ $\gamma$$\gamma$ in the $\sigma$-model}.
\newblock \emph{Il Nuovo Cimento A (1965-1970)}, 60\penalty0 (1):\penalty0
  47--61, 1969.

\bibitem[Son and Spivak(2013)]{Son:2013}
D.~T. Son and B.~Z. Spivak.
\newblock {Chiral anomaly and classical negative magnetoresistance of Weyl
  metals}.
\newblock \emph{Phys. Rev. B}, 88:\penalty0 104412, Sep 2013.
\newblock \doi{10.1103/PhysRevB.88.104412}.
\newblock URL \url{https://link.aps.org/doi/10.1103/PhysRevB.88.104412}.

\bibitem[Spivak and Andreev(2016)]{Spivak:2016}
B.~Z. Spivak and A.~V. Andreev.
\newblock {Magnetotransport phenomena related to the chiral anomaly in Weyl
  semimetals}.
\newblock \emph{Phys. Rev. B}, 93:\penalty0 085107, Feb 2016.
\newblock \doi{10.1103/PhysRevB.93.085107}.
\newblock URL \url{http://link.aps.org/doi/10.1103/PhysRevB.93.085107}.

\bibitem[Lucas et~al.(2016)Lucas, Davison, and Sachdev]{Lucas:2016}
Andrew Lucas, Richard~A Davison, and Subir Sachdev.
\newblock {Hydrodynamic theory of thermoelectric transport and negative
  magnetoresistance in Weyl semimetals}.
\newblock \emph{Proceedings of the National Academy of Sciences}, page
  201608881, 2016.

\bibitem[Hosur and Qi(2015)]{Hosur:2015}
Pavan Hosur and Xiao-Liang Qi.
\newblock {Tunable circular dichroism due to the chiral anomaly in Weyl
  semimetals}.
\newblock \emph{Physical Review B}, 91\penalty0 (8):\penalty0 081106, 2015.

\bibitem[Huang et~al.(2015)Huang, Zhao, Long, Wang, Chen, Yang, Liang, Xue,
  Weng, Fang, Dai, and Chen]{Huang:2015}
Xiaochun Huang, Lingxiao Zhao, Yujia Long, Peipei Wang, Dong Chen, Zhanhai
  Yang, Hui Liang, Mianqi Xue, Hongming Weng, Zhong Fang, Xi~Dai, and Genfu
  Chen.
\newblock {Observation of the Chiral-Anomaly-Induced Negative Magnetoresistance
  in 3D Weyl Semimetal TaAs}.
\newblock \emph{Phys. Rev. X}, 5:\penalty0 031023, Aug 2015.
\newblock \doi{10.1103/PhysRevX.5.031023}.
\newblock URL \url{http://link.aps.org/doi/10.1103/PhysRevX.5.031023}.

\bibitem[Gooth et~al.(2017)Gooth, Niemann, Meng, Grushin, Landsteiner,
  Gotsmann, Menges, Schmidt, Shekhar, S{\"u}{\ss}, et~al.]{Gooth:2017}
Johannes Gooth, Anna~C Niemann, Tobias Meng, Adolfo~G Grushin, Karl
  Landsteiner, Bernd Gotsmann, Fabian Menges, Marcus Schmidt, Chandra Shekhar,
  Vicky S{\"u}{\ss}, et~al.
\newblock {Experimental signatures of the mixed axial--gravitational anomaly in
  the Weyl semimetal NbP}.
\newblock \emph{Nature}, 547\penalty0 (7663):\penalty0 324, 2017.

\bibitem[Ma et~al.(2017)Ma, Xu, Chan, Zhang, Chang, Lin, Xie, Palacios, Lin,
  Jia, et~al.]{Ma:2017}
Qiong Ma, Su-Yang Xu, Ching-Kit Chan, Cheng-Long Zhang, Guoqing Chang, Yuxuan
  Lin, Weiwei Xie, Tom{\'a}s Palacios, Hsin Lin, Shuang Jia, et~al.
\newblock {Direct optical detection of Weyl fermion chirality in a topological
  semimetal}.
\newblock \emph{Nature Physics}, 13\penalty0 (9):\penalty0 842, 2017.

\bibitem[Bachmann et~al.(2017)Bachmann, Nair, Flicker, Ilan, Meng, Ghimire,
  Bauer, Ronning, Analytis, and Moll]{Bachmann:2017}
Maja~D Bachmann, Nityan Nair, Felix Flicker, Roni Ilan, Tobias Meng, Nirmal~J
  Ghimire, Eric~D Bauer, Filip Ronning, James~G Analytis, and Philip~JW Moll.
\newblock {Inducing superconductivity in Weyl semimetal microstructures by
  selective ion sputtering}.
\newblock \emph{Science Advances}, 3\penalty0 (5):\penalty0 e1602983, 2017.

\bibitem[Lv et~al.(2015{\natexlab{b}})Lv, Xu, Weng, Ma, Richard, Huang, Zhao,
  Chen, Matt, Bisti, et~al.]{Lv:2015b}
BQ~Lv, N~Xu, HM~Weng, JZ~Ma, P~Richard, XC~Huang, LX~Zhao, GF~Chen, CE~Matt,
  F~Bisti, et~al.
\newblock {Observation of Weyl nodes in TaAs}.
\newblock \emph{Nature Physics}, 11:\penalty0 724--727, 2015{\natexlab{b}}.

\bibitem[Zhang et~al.(2017{\natexlab{a}})Zhang, Yuan, Jiang, Tong, Zhang, Xie,
  and Jia]{Zhang:2015}
Cheng-Long Zhang, Zhujun Yuan, Qing-Dong Jiang, Bingbing Tong, Chi Zhang, X.~C.
  Xie, and Shuang Jia.
\newblock {Electron scattering in tantalum monoarsenide}.
\newblock \emph{Phys. Rev. B}, 95:\penalty0 085202, Feb 2017{\natexlab{a}}.
\newblock \doi{10.1103/PhysRevB.95.085202}.
\newblock URL \url{https://link.aps.org/doi/10.1103/PhysRevB.95.085202}.

\bibitem[Arnold et~al.(2016{\natexlab{a}})Arnold, Naumann, Wu, Sun, Schmidt,
  Borrmann, Felser, Yan, and Hassinger]{Arnold:2016}
F.~Arnold, M.~Naumann, S.-C. Wu, Y.~Sun, M.~Schmidt, H.~Borrmann, C.~Felser,
  B.~Yan, and E.~Hassinger.
\newblock {Chiral Weyl Pockets and Fermi Surface Topology of the Weyl Semimetal
  TaAs}.
\newblock \emph{Phys. Rev. Lett.}, 117:\penalty0 146401, Sep
  2016{\natexlab{a}}.
\newblock \doi{10.1103/PhysRevLett.117.146401}.
\newblock URL \url{https://link.aps.org/doi/10.1103/PhysRevLett.117.146401}.

\bibitem[Zhang et~al.(2016)Zhang, Xu, Belopolski, Yuan, Lin, Tong, Bian,
  Alidoust, Lee, Huang, Chang, Chang, Hsu, Jeng, Neupane, Sanchez, Zheng, Wang,
  Lin, Zhang, Lu, Shen, Neupert, Hasan, and Jia]{Zhang:2016}
Cheng-Long Zhang, Su-Yang Xu, Ilya Belopolski, Zhujun Yuan, Ziquan Lin,
  Bingbing Tong, Guang Bian, Nasser Alidoust, Chi-Cheng Lee, Shin-Ming Huang,
  Tay-Rong Chang, Guoqing Chang, Chuang-Han Hsu, Horng-Tay Jeng, Madhab
  Neupane, Daniel~S. Sanchez, Hao Zheng, Junfeng Wang, Hsin Lin, Chi Zhang,
  Hai-Zhou Lu, Shun-Qing Shen, Titus Neupert, M.~Zahid Hasan, and Shuang Jia.
\newblock {Signatures of the Adler-Bell-Jackiw chiral anomaly in a Weyl fermion
  semimetal}.
\newblock \emph{Nature Communications}, 7, Feb 2016.
\newblock ISSN 2041-1723.
\newblock \doi{10.1038/ncomms10735}.

\bibitem[Moll et~al.(2016)Moll, Potter, Nair, Ramshaw, Modic, Riggs, Zeng,
  Ghimire, Bauer, Kealhofer, Ronning, and Analytis]{Moll:2016}
Philip J.~W. Moll, Andrew~C. Potter, Nityan~L. Nair, B.~J. Ramshaw, K.~A.
  Modic, Scott Riggs, Bin Zeng, Nirmal~J. Ghimire, Eric~D. Bauer, Robert
  Kealhofer, Filip Ronning, and James~G. Analytis.
\newblock {Magnetic torque anomaly in the quantum limit of Weyl semimetals},
  journal = {Nature Communications}.
\newblock 7, Aug 2016.
\newblock ISSN 2041-1723.
\newblock \doi{10.1038/ncomms12492}.

\bibitem[Arnold et~al.(2016{\natexlab{b}})Arnold, Shekhar, Wu, Sun, dos Reis,
  Kumar, Naumann, Ajeesh, Schmidt, Grushin, et~al.]{Arnold:2016b}
Frank Arnold, Chandra Shekhar, Shu-Chun Wu, Yan Sun, Ricardo~Donizeth dos Reis,
  Nitesh Kumar, Marcel Naumann, Mukkattu~O Ajeesh, Marcus Schmidt, Adolfo~G
  Grushin, et~al.
\newblock {Negative magnetoresistance without well-defined chirality in the
  Weyl semimetal TaP}.
\newblock \emph{Nature communications}, 7, 2016{\natexlab{b}}.

\bibitem[Pippard(1989)]{Pippard:1989}
Alfred~Brian Pippard.
\newblock \emph{{Magnetoresistance in metals}}, volume~2.
\newblock Cambridge University Press, 1989.

\bibitem[dos Reis et~al.(2016)dos Reis, Ajeesh, Kumar, Arnold, Shekhar,
  Naumann, Schmidt, Nicklas, and Hassinger]{Reis:2016}
R.~D. dos Reis, M.~O. Ajeesh, N.~Kumar, F.~Arnold, C.~Shekhar, M.~Naumann,
  M.~Schmidt, M.~Nicklas, and E.~Hassinger.
\newblock {On the search for the chiral anomaly in Weyl semimetals: the
  negative longitudinal magnetoresistance}.
\newblock \emph{New Journal of Physics}, 18, Aug 11 2016.
\newblock ISSN 1367-2630.
\newblock \doi{10.1088/1367-2630/18/8/085006}.

\bibitem[Argyres and Adams(1956)]{Argyres:1956}
PN~Argyres and EN~Adams.
\newblock {Longitudinal magnetoresistance in the quantum limit}.
\newblock \emph{Physical Review}, 104\penalty0 (4):\penalty0 900, 1956.

\bibitem[Murzin(2000)]{Murzin:2000}
Sergei~Stanislavovich Murzin.
\newblock {Electron transport in the extreme quantum limit in applied magnetic
  field}.
\newblock \emph{Physics-Uspekhi}, 43\penalty0 (4):\penalty0 349--364, 2000.

\bibitem[Kim et~al.(2017)Kim, Ryoo, and Park]{Kim:2017}
Pilkwang Kim, Ji~Hoon Ryoo, and Cheol-Hwan Park.
\newblock {Breakdown of the Chiral Anomaly in Weyl Semimetals in a Strong
  Magnetic Field}.
\newblock \emph{Physical review letters}, 119\penalty0 (26):\penalty0 266401,
  2017.

\bibitem[Chan and Lee(2017)]{Chan:2017}
Ching-Kit Chan and Patrick~A. Lee.
\newblock {Emergence of gapped bulk and metallic side walls in the zeroth
  Landau level in Dirac and Weyl semimetals}.
\newblock \emph{Phys. Rev. B}, 96:\penalty0 195143, Nov 2017.
\newblock \doi{10.1103/PhysRevB.96.195143}.
\newblock URL \url{https://link.aps.org/doi/10.1103/PhysRevB.96.195143}.

\bibitem[Zhang et~al.(2017{\natexlab{b}})Zhang, Xu, Wang, Lin, Du, Guo, Lee,
  Lu, Feng, Huang, et~al.]{Zhang:2017c}
Cheng-Long Zhang, Su-Yang Xu, CM~Wang, Ziquan Lin, ZZ~Du, Cheng Guo, Chi-Cheng
  Lee, Hong Lu, Yiyang Feng, Shin-Ming Huang, et~al.
\newblock {Magnetic-tunnelling-induced Weyl node annihilation in TaP}.
\newblock \emph{Nature Physics}, 13\penalty0 (10):\penalty0 979--986,
  2017{\natexlab{b}}.

\bibitem[Popovi{\'c} et~al.(1997)Popovi{\'c}, Fowler, and
  Washburn]{Popovic:1997}
Dragana Popovi{\'c}, AB~Fowler, and S~Washburn.
\newblock {Metal-insulator transition in two dimensions: Effects of disorder
  and magnetic field}.
\newblock \emph{Physical review letters}, 79\penalty0 (8):\penalty0 1543, 1997.

\bibitem[Kravchenko et~al.(1994)Kravchenko, Kravchenko, Furneaux, Pudalov, and
  d’Iorio]{Kravchenko:1994}
Sergey~V Kravchenko, GV~Kravchenko, JE~Furneaux, Vladimir~M Pudalov, and
  M~d’Iorio.
\newblock {Possible metal-insulator transition at B= 0 in two dimensions}.
\newblock \emph{Physical Review B}, 50\penalty0 (11):\penalty0 8039, 1994.

\bibitem[Tanuma et~al.(1981)Tanuma, Inada, Furukawa, Takahashi, Iye, Onuki,
  Chikazumi, and Miura]{Tanuma:1981}
S~Tanuma, R~Inada, A~Furukawa, O~Takahashi, Y~Iye, Y~Onuki, S~Chikazumi, and
  N~Miura.
\newblock {Physics in High Magnetic Fields}.
\newblock \emph{Springer Series in Solid State Sciences}, 1981.

\bibitem[Fauqu\'e et~al.(2013)Fauqu\'e, LeBoeuf, Vignolle, Nardone, Proust, and
  Behnia]{Fauque:2013}
Beno\^{\i}t Fauqu\'e, David LeBoeuf, Baptiste Vignolle, Marc Nardone, Cyril
  Proust, and Kamran Behnia.
\newblock {Two Phase Transitions Induced by a Magnetic Field in Graphite}.
\newblock \emph{Phys. Rev. Lett.}, 110:\penalty0 266601, Jun 2013.
\newblock \doi{10.1103/PhysRevLett.110.266601}.
\newblock URL \url{http://link.aps.org/doi/10.1103/PhysRevLett.110.266601}.

\bibitem[Zhu et~al.(2017)Zhu, McDonald, Shekhter, Ramshaw, Modic, Balakirev,
  and Harrison]{Zhu:2017}
Zengwei Zhu, RD~McDonald, A~Shekhter, BJ~Ramshaw, KA~Modic, FF~Balakirev, and
  N~Harrison.
\newblock {Magnetic field tuning of an excitonic insulator between the weak and
  strong coupling regimes in quantum limit graphite}.
\newblock \emph{Scientific reports}, 7\penalty0 (1):\penalty0 1733, 2017.

\bibitem[Pippard(1955)]{Pippard:1955}
AB~Pippard.
\newblock {Ultrasonic Attenuation in Metals}.
\newblock \emph{Philosophical Magazine}, 46\penalty0 (381):\penalty0
  1104--1114, 1955.
\newblock ISSN 0031-8086.

\bibitem[LeBoeuf et~al.(2017)LeBoeuf, Rischau, Seyfarth, K{\"u}chler, Berben,
  Wiedmann, Tabis, Frachet, Behnia, and Fauqu{\'e}]{LeBoeuf:2017}
David LeBoeuf, CW~Rischau, Gabriel Seyfarth, R~K{\"u}chler, Martin Berben,
  Steffen Wiedmann, Wojciech Tabis, Medhi Frachet, Kamran Behnia, and Benoit
  Fauqu{\'e}.
\newblock {Thermodynamic signatures of the field-induced states of graphite}.
\newblock \emph{Nature communications}, 8\penalty0 (1):\penalty0 1337, 2017.

\bibitem[Spivak and Kivelson(2006)]{Spivak:2006}
Boris Spivak and Steven~A Kivelson.
\newblock {Transport in two dimensional electronic micro-emulsions}.
\newblock \emph{Annals of Physics}, 321\penalty0 (9):\penalty0 2071--2115,
  2006.

\bibitem[Paalanen et~al.(1992)Paalanen, Willett, Littlewood, Ruel, West,
  Pfeiffer, and Bishop]{Paalanen:1992}
M.~A. Paalanen, R.~L. Willett, P.~B. Littlewood, R.~R. Ruel, K.~W. West, L.~N.
  Pfeiffer, and D.~J. Bishop.
\newblock {RF conductivity of a two-dimensional electron system at small
  Landau-level filling factors}.
\newblock \emph{Phys. Rev. B}, 45:\penalty0 11342--11345, May 1992.
\newblock \doi{10.1103/PhysRevB.45.11342}.
\newblock URL \url{https://link.aps.org/doi/10.1103/PhysRevB.45.11342}.

\bibitem[Shekhter et~al.(2013)Shekhter, Ramshaw, Liang, Hardy, Bonn, Balakirev,
  McDonald, Betts, Riggs, and Migliori]{Shekhter:2013}
Arkady Shekhter, BJ~Ramshaw, Ruixing Liang, WN~Hardy, DA~Bonn, Fedor~F
  Balakirev, Ross~D McDonald, Jon~B Betts, Scott~C Riggs, and Albert Migliori.
\newblock {Bounding the pseudogap with a line of phase transitions in
  YBa2Cu3O$_{6+\delta}$}.
\newblock \emph{Nature}, 498\penalty0 (7452):\penalty0 75--77, 2013.

\bibitem[Field et~al.(1986)Field, Reich, Shivaram, Rosenbaum, Nelson, and
  Littlewood]{Field:1986}
Stuart~B. Field, D.~H. Reich, B.~S. Shivaram, T.~F. Rosenbaum, D.~A. Nelson,
  and P.~B. Littlewood.
\newblock {Evidence for depinning of a Wigner crystal in Hg-Cd-Te}.
\newblock \emph{Phys. Rev. B}, 33:\penalty0 5082--5085, Apr 1986.
\newblock \doi{10.1103/PhysRevB.33.5082}.
\newblock URL \url{https://link.aps.org/doi/10.1103/PhysRevB.33.5082}.

\bibitem[Shoenberg(1984)]{Shoenberg:1984}
D.~Shoenberg.
\newblock \emph{{Magnetic oscillations in metals}}.
\newblock Cambridge monographs on physics. Cambridge University Press, 1984.
\newblock ISBN 9780521224802.

\bibitem[Ramshaw et~al.(2011)Ramshaw, Vignolle, Day, Liang, Hardy, Proust, and
  Bonn]{Ramshaw:2011}
BJ~Ramshaw, Baptiste Vignolle, James Day, Ruixing Liang, WN~Hardy, Cyril
  Proust, and DA~Bonn.
\newblock {Angle dependence of quantum oscillations in YBa2Cu3O6. 59 shows
  free-spin behaviour of quasiparticles}.
\newblock \emph{Nature Physics}, 7\penalty0 (3):\penalty0 234--238, 2011.

\bibitem[Brugger(1965)]{Brugger:1965}
K~Brugger.
\newblock {Pure modes for elastic waves in crystals}.
\newblock \emph{Journal of Applied Physics}, 36\penalty0 (3):\penalty0
  759--768, 1965.

\bibitem[Suslov et~al.(2006)Suslov, Sarma, Ketterson, Balakirev, Migliori, and
  Lacerda]{Suslov:2006}
A~Suslov, Bimal~K Sarma, JB~Ketterson, F~Balakirev, A~Migliori, and A~Lacerda.
\newblock {Ultrasonic instrumentation for measurements in high magnetic fields.
  II. Pulsed magnetic fields}.
\newblock \emph{Review of scientific instruments}, 77\penalty0 (3):\penalty0
  035105, 2006.

\bibitem[Turner et~al.(2012)Turner, Zhang, Mong, and Vishwanath]{Turner:2012}
Ari~M. Turner, Yi~Zhang, Roger S.~K. Mong, and Ashvin Vishwanath.
\newblock {Quantized response and topology of magnetic insulators with
  inversion symmetry}.
\newblock \emph{Phys. Rev. B}, 85:\penalty0 165120, Apr 2012.
\newblock \doi{10.1103/PhysRevB.85.165120}.
\newblock URL \url{https://link.aps.org/doi/10.1103/PhysRevB.85.165120}.

\bibitem[Hu et~al.(2016)Hu, Liu, Graf, Radmanesh, Adams, Chuang, Wang,
  Chiorescu, Wei, Spinu, et~al.]{Hu:2016}
Jin Hu, JY~Liu, David Graf, SMA Radmanesh, DJ~Adams, Alyssa Chuang, Yu~Wang,
  Irinel Chiorescu, Jiang Wei, Leonard Spinu, et~al.
\newblock {$\pi$ Berry phase and Zeeman splitting of Weyl semimetal TaP}.
\newblock \emph{Scientific reports}, 6:\penalty0 18674, 2016.

\end{thebibliography}
\end{document}